\documentclass[aps,twocolumn,groupedaddress,floatfix]{revtex4}
\usepackage[colorlinks=true,citecolor=red,filecolor=green,linkcolor=blue,pdfnewwindow=true]{hyperref}
\usepackage{amsmath} 
\usepackage[version=4]{mhchem}
\usepackage{color}
\usepackage{graphicx}
\usepackage{makeidx}
\usepackage{bm}
\usepackage[title]{appendix} 
\usepackage{tabu}
\usepackage{float}
\usepackage{caption}
\usepackage{hyperref}
\usepackage[normalem]{ulem}
\usepackage{natbib}
\usepackage[dvipsnames]{xcolor}

\hypersetup{urlcolor=blue}

\def\la{\langle}
\def\ra{\rangle}

\makeatletter
 
\newcommand{\Rmnum}[1]{\expandafter\@slowromancap\romannumeral #1@}
\makeatother
\begin{document}

\title{Analysis of the structural complexity of Crab nebula observed at\\ 
radio frequency using a multifractal approach}

\author{Athokpam Langlen Chanu$^{1,4}$}                                    
\email{athokpam.chanu@apctp.org (Corresponding\\
author)}
\author{Pravabati Chingangbam$^{2}$}                                 
\email{prava@iiap.res.in}

\author{\\ Fazlu Rahman$^{2,3}$}  
\email{fazlu.rahman@iiap.res.in}

\author{R.~K.~Brojen Singh$^4$}
\email{brojen@jnu.ac.in}

\author{Preeti Kharb$^5$}
\email{kharb@ncra.tifr.res.in}
\affiliation{$^1$ Asia Pacific Center for Theoretical Physics, Pohang, 37673, Republic of Korea\\
$^2$ Indian Institute of Astrophysics, Koramangala II Block,  
  Bangalore 560034, India\\
        $^3$ Department of Physics, Pondicherry University, R.V. Nagar, Kalapet, 605014, Puducherry, India\\
        $^4$ School of Computational and Integrative Sciences, Jawaharlal Nehru~University, New Delhi 110067, India\\
        $^5$ National Centre for Radio Astrophysics - Tata Institute of Fundamental Research, Savitribai Phule Pune University Campus, Pune 411007, Maharashtra, India}

\begin{abstract}
\noindent The Crab Nebula is an astrophysical system that exhibits complex morphological patterns at different observing frequencies. We carry out a systematic investigation of the structural complexity of the nebula using publicly available imaging data at radio frequency. For the analysis, we use the well-known multifractal detrended fluctuation analysis (MFDFA) in two dimensions. We find that radio data exhibit long-range correlations, as expected from the underlying physics of the supernova explosion and evolution. The correlations follow a power-law scaling with length scales. The structural complexity is found to be multifractal in nature, as evidenced by the dependence of the generalized Hurst exponent on the order of the moments of the detrended fluctuation function. By repeating the analysis on shuffled data, we further probe the origin of the multifractality in the radio imaging data. For the radio data, we find that the probability density function (PDF) is close to a Gaussian form. Hence, the multifractal behavior is due to the differing nature of long-range correlations of the large and small detrended fluctuation field values. We investigate the multifractal parameters across different partitions of the radio image and find that the structures across the image are highly heterogeneous, making the Crab Nebula a structurally complex astrophysical system. Our analysis thus provides a fresh perspective on the morphology of the Crab Nebula from a complexity science viewpoint.\\

{\noindent}
\end{abstract}

\maketitle


\section{Introduction}
\noindent Natural systems are generally complex, consisting of a large number of interlinked/integrated components of heterogenous properties \cite{Mitchell}. The individual components of such complex systems have non-linear character \cite{kwapien2012physical, gershenson2005can}. Their interaction can give rise to emergent behaviors at different scales. An important property of complex systems is self-organization \cite{kauffman1993origins}, which can be captured by fractals or multifractals. The notion of multifractality originated in the study of turbulence in fluid mechanics \cite{novikov1971intermittency, mandelbrot1999intermittent}. Classically, multifractals are analyzed using two approaches: structure-function \cite{van1970structure} and partition function \cite{benzi1984multifractal}. While the structure-function approach is based on moments of increment distribution, the partition-function approach mainly adopts the box-counting formalism to determine the multifractal spectrum ($\alpha,f$). Other approaches to analyzing multifractals in time-series data include the wavelet transform approach, fluctuation analysis approach, and detrended fluctuation approach. We refer interested readers to the excellent review \cite{jiang2019multifractal} for details of these approaches. The well-established multifractal formalism is used across a wide variety of fields. It is used to study long-term persistence in river precipitation (earth sciences) \cite{kantelhardt2006long}, dynamics of stock markets (econophysics) \cite{jiang2019multifractal}, anomalies in the dynamics of brain (neuroscience) \cite{islam2013multifractal}, social media activities (social dynamics) \cite{oh2022multifractal} and species abundance distributions (ecology) \cite{borda2002species}. Kantelhardt \textit{et al.} \cite{Kantelhardt_2002} developed a numerical method known as the multifractal detrended fluctuation analysis (MFDFA) to calculate various multifractal parameters for characterizing multifractal scaling properties and detecting long-range correlations in noisy, non-stationary time series data of complex systems. The method is based on a generalization of the detrended fluctuation analysis (DFA) initially given by Peng \textit{et al} \cite{Peng_1994}. Gu and Zhou \cite{Gu_2006} extended the conventional MFDFA to the two-dimensional multifractal detrended fluctuation analysis (2D-MFDFA) to analyze higher dimensional fractals and multifractals.

The multifractal theoretical framework has long been studied by theoretical physicists and mathematicians. The multifractal spectrum, $f(\alpha$), also known as \textit{H$\ddot{o}$lder spectrum}, associates with each positive $\alpha$ the Hausdorff dimension of the set \cite{rand1989singularity}. For instance, in the case of fully developed turbulence, the H$\ddot{o}$lder spectrum is determined for the velocity of the fluid \cite{olsen1995multifractal}. The formalism of fractal or multifractal analysis \cite{salat2017multifractal} runs parallel to mathematical frameworks, such as multifractal decomposition of Moran fractals \cite{cawley1992multifractal}, and digraph recursive fractals \cite{edgar1992multifractal}, maximal measure \cite{lopes1989dimension}, discrete measures \cite{aversa1990multifractal}, the theory of large deviations \cite{harte2001multifractals}, Minkowski measurability \cite{carfi2013fractal}, self-similar functions \cite{jaffard1997multifractal1,jaffard1997multifractal2}, etc. The present paper aims to introduce and explore a fractal theory and a multifractal analysis in the context of a complex astrophysical system, namely, the Crab Nebula. Astrophysics provides a multitude of examples of complex systems owing to the different physical interactions and evolution time scales involved. Examples of complex astrophysical systems include turbulent plasma, such as stellar flares and accretion discs, supernova remnants, spatial structures in the interstellar medium, galaxies, and the large-scale structure of the Universe \cite{shore2003galaxies,regev2006chaos,aschwanden2011self}. To measure the complexity of their dynamics or morphological structures from observational data is an intriguing and active research topic. Supernova remnants are particularly interesting for questions related to complexity because of their complicated morphological patterns that evolve with time. In this paper, we focus our attention on the Crab Nebula, a well-known remnant of a supernova recorded in 1054 AD~\cite{Duyvendak:1942}. It is located about 6500 light-years from the earth~\cite{Trimble:1973} and has been well-studied using multi-wavelength observations from radio frequencies to gamma rays (see reviews \cite{Hester:2008,Buehler:2014}). We choose to study the Crab Nebula from the viewpoint of complex systems because of its historical importance and availability of imaging data with good spatial resolution in several frequency bands. Several studies have been carried out to investigate the morphological variation of the Crab Nebula across different frequencies (see e.g.~\cite{Dubner:2017, Yeung_2019,Fang_2019}). However, a quantitative analysis of the morphological complexity of the Crab Nebula is lacking in the literature.

The intricate morphology of the Crab Nebula that can be easily discerned by the eye from high-resolution images naturally leads to the question of its structural complexity.  A complex systems approach to the Crab Nebula views the physical system as one system of inter-connected components with their organization exhibiting emergent behaviors instead of studying the dynamics of the individual components. To investigate the structural complexity of the complex system of the Crab Nebula, we adopt a multi-scale multifractal approach. A multifractal approach is very well-suited to study and analyze the structural properties of the Crab Nebula since it provides a framework for detecting and identifying complex local structures and describing local singularities. In astrophysics, multifractal formalism is applied to study the large scale structure of the universe~\cite{Jones:1988, gaite2019fractal}, the behavior of galaxy clustering on large scales \cite{pan2000large}, the interstellar medium \cite{Elia_2018}, Gamma-ray bursts \cite{Meredith1995,PhysRevA.40.5284}, flux correlations of pulsars \cite{Eghdami_2018}, flux variability in the quasar 3C~273 \cite{Belete_2018}, gravitational wave signals detected by LIGO \cite{de_Freitas_2018}, to name a few. The well-established one-dimensional MFDFA is used to study sunspot number fluctuations \cite{Movahed_2006}, investigate the statistical properties of the cosmic microwave background radiation \cite{Movahed_2011}, and analyze the pseudorapidity distribution of ring-like events of charged mesons \cite{Haldar}. We systematically investigate the structural complexity of the Crab Nebula observed at radio frequency using publicly available data with the 2D-MFDFA method. Our findings provide the local scaling behavior, nature of correlations, the origin of multifractality, heterogeneous properties, and organization in the structural system of the Crab Nebula. Our results deepen our understanding of the multi-scale physics operating in the Crab Nebula. 

This paper is organized as follows. Section~\ref{sec:2} outlines the methodology used in our analysis. We briefly introduce the concept of multifractality and describe the 2D-MFDFA algorithm and the associated physical interpretations. Section~\ref{sec:3} describes the physics of the Crab Nebula and the imaging data used in our analysis. In section~\ref{sec:4}, we present our main analysis and results. We conclude with a summary and discuss the implications of our results in section~\ref{sec:5}.
\section{Methodology}
\label{sec:2}
In this section, we briefly review multifractality and complexity measures, then present the 2D-MFDFA algorithm and its associated physical interpretations.
\subsection{Complexity and multifractality}
{\noindent}
Given $y(0)=0$, a random process $\{y(t)\}$ is said to be a self-affine process if it obeys the following scaling relation \cite{RePEc:cwl:cwldpp:1164}
\begin{align}
	\frac{y(ct)}{y(t)}=c^{D} \ ; \ \forall \ c>0, \label{eq:selfaffine1}
\end{align}
where $c$ is the scale factor. The scaling exponent $D>0$ represents the self-affine process's fractal or self-similarity dimension. A uni-fractal or uniscaling process is characterized by a single scaling law at any scale. Data generated by various complex systems exhibit fluctuations on a broad range of scales. Such fluctuations often follow a scaling relation of the type \eqref{eq:selfaffine1} over several orders of magnitude in both equilibrium and non-equilibrium situations \cite{Kantelhardt_2002}.

\noindent The theory of multifractals studies a more general relation of the type 
\begin{equation}
 \frac{y(ct)}{y(t)}= \Gamma(c)\ ; \forall t, \ 0<c\leq 1, \label{eq:selfaffine2}
 \end{equation}
where $y(t)$ and $\Gamma(c)$ are independent random functions. If the random process $\{y(t)\}$ is multifractal, then the scaling function $\Gamma(c)$ satisfies \cite{RePEc:cwl:cwldpp:1164}
\begin{align}
 \Gamma(c_1c_2\dots c_n)&=\Gamma_1(c_1)\ \Gamma_2(c_2)\dots \Gamma_n(c_n);  \nonumber \\
 & \ 0 <c_1,c_2,\dots,c_n\leq 1 \label{eq:selfaffine3}
  \end{align}
where $\Gamma_1,\ \Gamma_2, \ \dots \Gamma_n$ are $n$ independent copies of $\Gamma$ for various local scales $c_1, c_2, \dots, c_n$.
Using eq.~\eqref{eq:selfaffine1} and \eqref{eq:selfaffine2}, each local scale $c_{\kappa}$ of eq.~\eqref{eq:selfaffine3} has the local fractal dimension $D_{\kappa}$ and hence the local scaling function follows the relation
 \begin{equation}
  \Gamma_{\kappa}(c_{\kappa})\sim c_{\kappa}^{D_{\kappa}}. \label{eq:selfaffine4}
 \end{equation} 
The multifractal process becomes a monofractal when $c_1=c_2=\dots =c_n=c$ in eq.~\eqref{eq:selfaffine3}. Using eq.~\eqref{eq:selfaffine4}, we can write the scaling function $\Gamma$ of eq.~\eqref{eq:selfaffine3} as $\Gamma=c^{D1+D_2+\dots+D_n}=c^D$, where $D=D_1+D_2+...+D_n$, which is exactly eq.~\eqref{eq:selfaffine1}. Thus, for a uni-fractal system, the scaling function $\Gamma$ characterizes a homogenous fractal structure/behavior with a single scaling exponent $D$ at all scales $c$. Multifractality, on the other hand, can provide a richer variety of structures/behaviors. A multifractal analysis thus can be used to describe the fluctuations-driven local patterns/clusters of a data field represented by a set of scaling exponents corresponding to the local patterns.
\subsection{Review of two-dimensional Multifractal Detrended Fluctuation Analysis (2D-MFDFA)}
We briefly review here the algorithm and physical interpretations of the two-dimensional MFDFA method, following~\cite{Kantelhardt_2002,Gu_2006}
\subsubsection{2D-MFDFA algorithm}
Consider a two-dimensional compact space $S$, which is a subset of flat two-dimensional space. For our purpose, $S$ is a rectangular region that is subdivided into $M\times N$ equal-area square pixels. Let $d$ denote the data field, a function on $S$. The $d$ is thus a two-dimensional $M\times N$ array.

The algorithm for the 2D-MFDFA involves the following steps. 
\begin{enumerate}
\item Partition $S$ into $M_s\times N_s$ disjoint square subsets, with each subset containing $s^2$ pixels. We refer to $s$ as the {\em scale size}. Let us call each subset a {\em superpixel}. Let each superpixel be  indexed by $I,J$, such that $I=1,2,\ldots,M_s$ and $J=1,2,\ldots,N_s$.  We assume that $M, N$ are multiples of $s$ so that $M_s=M/s$ and $N_s=N/s$ are positive integers. 
We then carry out the partitioning for different values of $s$. For a given $s$, all superpixels contain the same number of pixels.
\item Let $d_{I,J}$ denote the $s\times s$ data array within each superpixel $I,J$. The {\em cumulative sum} $U_{I,J}$ of the data  
for each superpixel is defined by 
\begin{equation}
U_{I,J}(i,j)=\sum_{k_1=1}^{i} \sum_{k_2=1}^{j}  d_{I,J}(k_1,k_2),
\end{equation}
where $i,j=1,\ldots,s$ are indices for the pixels within the superpixel.
\item To subtract the trend in each superpixel, we first fit $U_{I, J}(i, j)$ with a  linear bivariate polynomial function $U_{I, J}^{\rm fit}(i, j)$, given by 
\begin{equation}
    U_{I,J}^{\rm fit}(i, j)=Ai+Bj+C,
\end{equation}
where $A, B, C$ are parameters determined by the fitting process. In general, one can also use higher-order polynomials for the fitting function. The detrended fluctuation function that is defined in the steps below will differ based on the order of the fitting function. We have chosen a linear form since it is the simplest. Even though the numerical values of our derived quantities are affected by the order of the fitting polynomial, the overall conclusions of our results remain the same. We then calculate the residual
\begin{equation}
\Delta U_{I,J}(i,j)=U_{I,J}(i, j)-U_{I,J}^{\rm fit}(i, j).
\end{equation}
Then, for a given $s$, the detrended fluctuation function of each superpixel, which we denote by $G(I, J,s)$, is calculated as
\begin{equation}
G^2(I,J,s)=\frac{1}{s^2}\sum_{i=1}^{s} \sum_{j=1}^{s} \Delta U_{I,J}(i,j)^2.
\label{eq:G}
\end{equation}

\item Let $q$ be a non-zero real number. The {\em detrended fluctuation function} of order $q$, denoted by $F_q(s)$, is calculated by averaging over all the  superpixels as
\begin{equation}
F_q(s)=\left \{\frac{1}{M_sN_s} \sum_{I=1}^{M_s} \sum_{J=1}^{N_s}[G(I,J,s)]^q\right \}^{1/q}.
\label{eq:ffunc}
\end{equation}
For positive integer values of $q$, this expression reduces to the usual definition of the $q^{th}$ order moment of $G(I, J,s)$. 

For the case of $q=0$, eq.~\eqref{eq:ffunc} cannot be used since it diverges. Instead, we use the following expression for $F_{0}(s)$
\begin{align}
F_0(s)=\exp\left \{\frac{1}{M_sN_s} \sum_{I=1}^{M_s} \sum_{J=1}^{N_s}\ln[G(I,J,s)]\right \}.
\label{eq:Fq0}
\end{align}
Since we use the positive square root of eq.~\eqref{eq:G},  $F_q(s)$  is always real.

\item For different values of $q$, the scaling relation between $F_q(s)$ and the scale size $s$ can be expressed as
\begin{equation}
F_q(s)\sim s^{h(q)}, \label{eq:scaling}
\end{equation}
where the scaling exponent $h(q)$ is the {\em generalized Hurst exponent}. The $h(q)$ is a measure of self-similarity symmetry and correlation present in the data field. We restrict our attention here to data for which $F_q(s)$ are increasing functions of $s$, or positive values of $h(q)$. This means that the data of interest contains {\em positive} correlations. 
\end{enumerate}

Mathematically, there is a-priori no restriction on the range of $q$ values. However, the physical information that can be obtained may saturate beyond some range of $q$ values,  such as what we will find when $h(q)$ asymptotes to constant values as $q$ increases. 

\subsubsection{Physical interpretations}
\label{sec:sec2B2}

As described above the 2D-MFDFA algorithm determines the scaling relation between $F_q(s)$ and $s$, from which we can calculate $h(q)$. For a general data field, $h$ can depend on both $q$ and $s$.  From the behaviour of $h(q)$, we can infer the following:  
\begin{itemize}
\item  If the data contains long-range power-law correlations, then the dependence of $F_q(s)$ on $s$ has a power-law form, and $h(q)$ is independent of $s$.     

\item If $h(q)$ remains the same for varying  $q$, then the data field has {\em monofractal} scaling property.  This means that the data or physical system contains one structural element or pattern that is invariant under size scale transformations. Then only one scaling exponent, namely $h$ describes the self-similar scaling behavior of the system.
\item If, however, $h$ has a dependence on $q$, then the data has {\em multifractal} scaling behaviour. This implies that the data contains more than one structural element, which is invariant under size scale transformations. 
This further shows that small and large fluctuations scale differently with $s$. 
\item 
For positive values of $q$, the dominant contributions to the sum on the right-hand side of eq.~\ref{eq:ffunc} will come from superpixels containing large fluctuations (or equivalently large deviations from the fitting function $U_{I, J}^{\rm fit}$ given in step 3 in the previous subsection). So, for $q>0$, $h(q)$ effectively describes the scaling behavior of large fluctuations. In contrast,  for negative values of $q$, the dominant contribution for the averaging in eq.~\ref{eq:ffunc} will come from superpixels containing small values of fluctuations. Hence in the region $q<0$, $h(q)$  effectively describes the scaling behavior of small fluctuations.
\item Furthermore, given some data having multifractal properties, the value of $h(q)$ is usually larger for $q<0$, compared to $q>0$. This can be explained by the fact that the contributions to the averaging in eq.~\ref{eq:ffunc} from small fluctuations typically vary more across the scale $s$, compared to large fluctuations. 
\end{itemize}

We next relate $h(q)$  to other measures of complexity used in the literature.  From the standard partition function-based multifractal formalism, the {\em classical multifractal scaling exponent} $\tau$ is related to $h(q)$ as \cite{Kantelhardt_2002}
\begin{equation}
\tau(q)=qh(q)-D_f, \label{eq:tau}
\end{equation}
where $D_f$ is the fractal dimension of the geometric support of the object. We take $D_f=2$ for a two-dimensional (2D) image \cite{Gu_2006}. If $h$ is independent of $q$, then $\tau$ is a linear function of $q$. Therefore, a nonlinear shape of $\tau(q)$ is an indication of the multifractality of the system.

The {\em H$\ddot{o}$lder exponent} (also known as {\em singularity strength}), denoted by $\alpha$, is calculated by taking the Legendre transformation of eq.~\eqref{eq:tau} as
\begin{equation}
\alpha(q)=\tau '(q), \label{eq:alpha}
\end{equation}
where the prime denotes derivative with respect to $q$. 
We get
\begin{equation}
\alpha(q)= h(q)+qh'(q).
\label{eq:aq}
\end{equation}
$\alpha$ is a measure of the shape of $h(q)$ since it is a  function of $h'(q)$. Different values of $\alpha$ characterise different parts of the system roughly since small, intermediate or large values of the fluctuation field will contribute to different values of $\alpha$. From eq.~\ref{eq:aq} we can see that $\alpha$ can take both positive and negative real values depending on the sign of the second term and its relative value with respect to the first term. Note that for a monofractal system, we have $h'(q) = 0$ since $h$ is a constant, and as a consequence, $\alpha$ has only one value given by $\alpha = h$. Therefore, a variation of $\alpha$ with $q$ signifies multifractality, and the range over which it varies quantifies the strength of multifractality.

The last quantity we use is the {\em multifractal spectrum} (also called the {\em singularity spectrum}) $f(\alpha)$, defined by
\begin{equation}
 f(\alpha)=q\alpha-\tau(q). \label{eq:falpha}
\end{equation}
This quantity is useful in extracting information about the symmetry between small and large fluctuations. 

\begin{figure}
Example A \hskip 3cm Example B
\includegraphics[height=9cm,width=8.5cm]{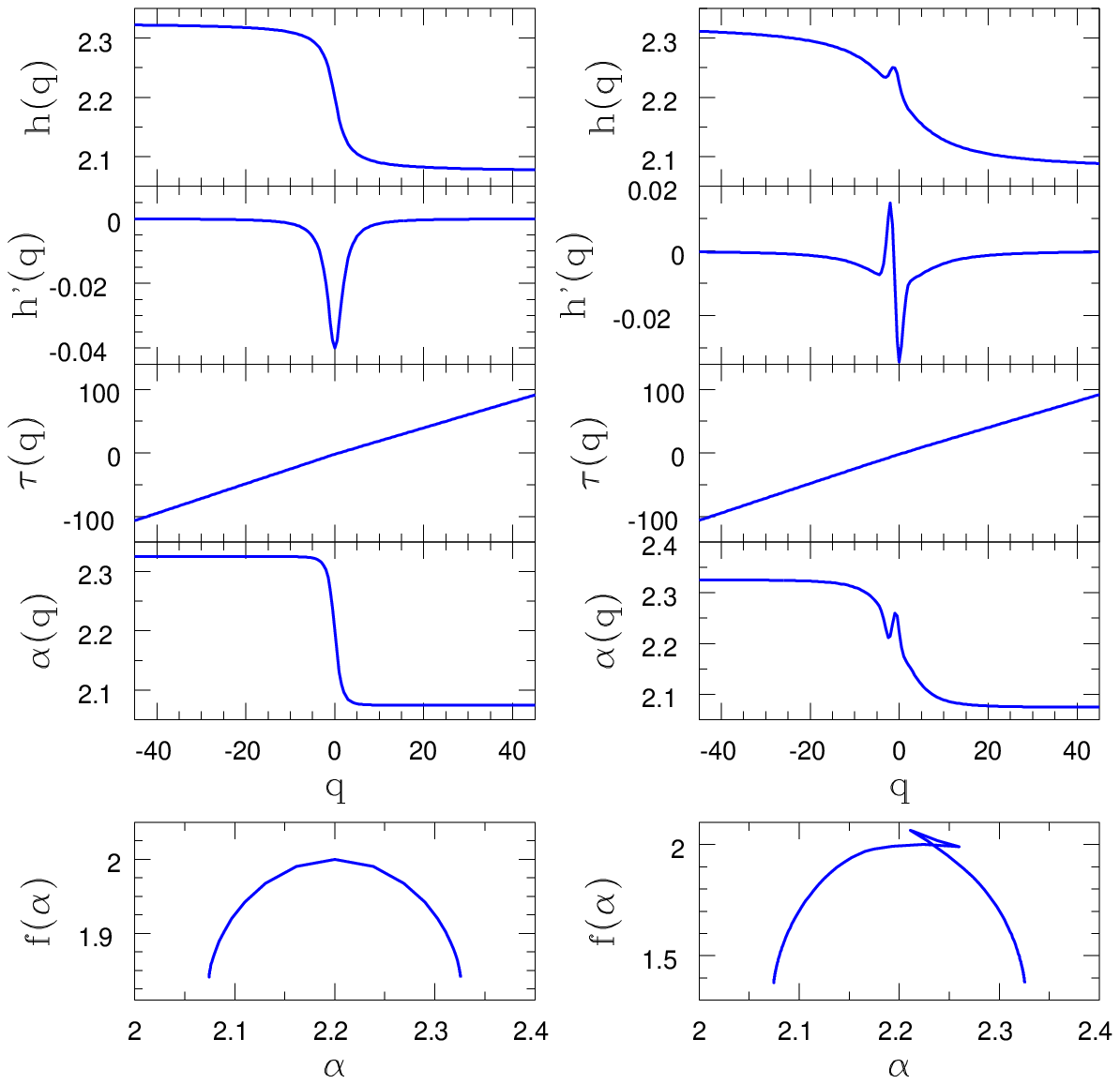}
  \caption{$h,\ h', \tau,\ \alpha$ are plotted versus $q$, and $f$ is plotted versus $\alpha$, for the toy examples given by eqs.~\ref{eq:toy1} and \ref{eq:toy2}.}
  \label{fig:exponents}
  \end{figure}

In order to gain an intuitive understanding of the quantities we have defined so far, let us consider two toy examples of multifractal complex systems whose $h(q)$ are given below.
\begin{description}
\item[Example A.] Let $h(q)$ be given by
\begin{equation}
h(q)=2.2-0.08\,{\rm tan}^{-1}\left(\frac{q}{2} \right).
\label{eq:toy1}
\end{equation}
\item[Example B.] Let $h(q)$ be given by 
\begin{equation}
h(q)=2.2-0.08\,{\rm tan}^{-1}\left( \frac{q}{2} \right) + 0.04\, e^{- (q+1)^2/2}. 
\label{eq:toy2}
\end{equation}
\end{description}
These examples have been chosen so as to capture the essence of our results in section~\ref{sec:4}. Figure~\ref{fig:exponents} shows $h$, $h'$, $\tau$ and $\alpha$ versus $q$, and $f$ versus $\alpha$ for the two examples. We have chosen $D_f=2$ for both examples.  If $h(q)$ is a well-behaved monotonic function of $q$, such as in example A, then all the other quantities are single-valued. However, if $h(q)$ is not monotonic, such as in example B, then $h'$ and $\alpha$ become multi-valued. As a consequence, the multifractal spectrum $f(\alpha)$ is not a well-behaved function.

In case, $f(\alpha)$ has a well-defined peak located at $\alpha=\alpha_0$, then we can quantify its symmetry about $\alpha_0$.  Let $\alpha_{\rm min}$ be  the smallest value of $\alpha$, and $\alpha_{\rm max}$ denote the value of $\alpha$ where $f(\alpha_{\rm max}) = f(\alpha_{\rm min})$. Then, let $\chi$ denote the skewness parameter, given by
\begin{equation}
    \chi \equiv 
    \frac{|\alpha_{\rm max}-\alpha_0|}{|\alpha_{\rm min}-\alpha_0|}.
    \label{eq:skewness}
\end{equation}
The $f(\alpha)$ is right-skewed if $\chi>1$, left-skewed if $\chi<1$, and  symmetrical if $\chi=1$. If $\chi<1$, then it shows that the scaling of the system is dominated by large fluctuations (smaller generalized Hurst exponents). However, if $\chi>1$, then it shows the dominance of scaling by small fluctuations (higher generalized Hurst exponents).

\section{The Crab nebula - physical processes and observed data}
\label{sec:3}

 In this section, we present a brief description of the physical processes that are relevant to understanding the Crab Nebula. Then we give details of the imaging data that we use for our analysis.
 
\subsection{Physics of the Crab Nebula}

The Crab Nebula belongs to the class of supernova remnants known as pulsar wind nebulae (PWN). Such a nebula inherits its morphology and spectrum from the following factors. 
\begin{enumerate}
\item During the supernova explosion of the progenitor star, the stellar material gets ejected (known as {\em ejecta}). The ejecta sweeps up the interstellar medium, creates a heated shock front, slows down, and eventually freely expands into the interstellar regions beyond the confines of the nebula.  
\item The collapse of the core of the progenitor star results in a rotating magnetized neutron star, known as the Crab pulsar. The magnetic field causes the charged ejecta particles to accelerate and emit synchrotron radiation. 

\item The Crab pulsar spins at the rate of 33 ms~\cite{Rosat:1995}. The pulsar emits non-thermal plasma, known as pulsar wind, which gets accelerated to high velocities by the magnetic field. The plasma pushes into the surrounding ejecta in the nebula and undergoes turbulent mixing. This leads to Rayleigh-Taylor instabilities~\cite{Chevalier:1975} that give shape to the Crab Nebula's filamentary network and wispy structures. 
\end{enumerate}

\begin{figure}
\includegraphics[height=6cm,width=7cm]{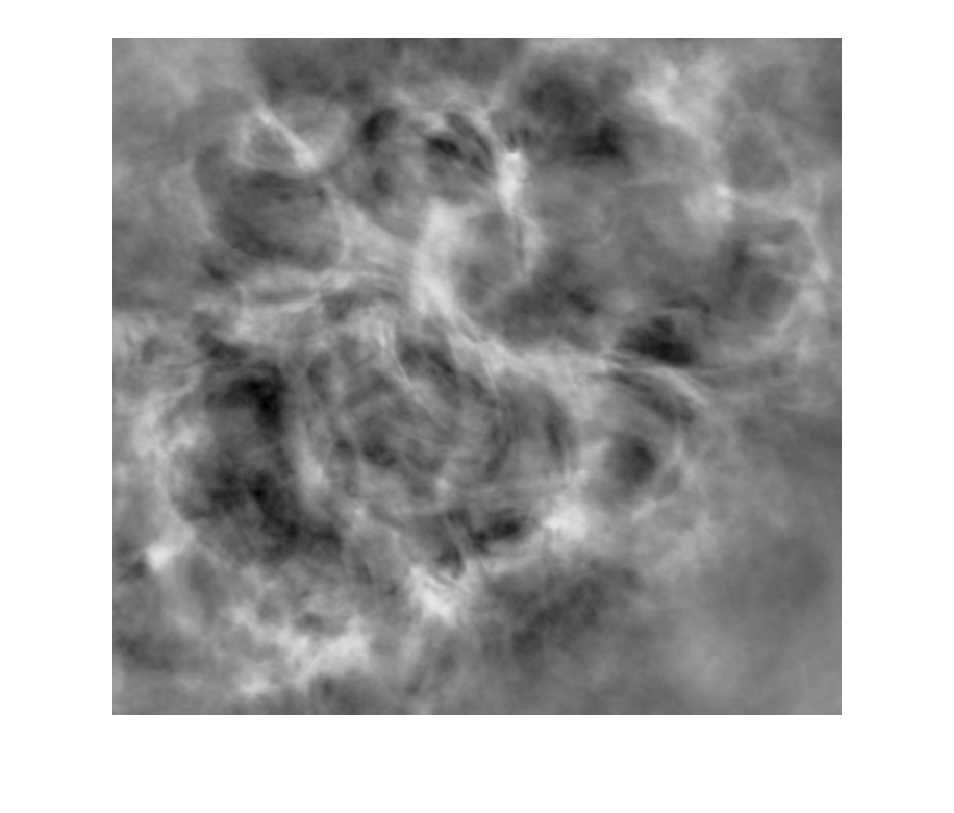}
  \caption{VLA image of Crab nebula at 4.76 GHz.
  }
  \label{fig:fig1}
  \end{figure}
  
The Crab Nebula appears like a bubble whose shell expands radially over time as the ejected material moves outward, while the interior structure consists of an expanding network of filaments and wispy structures when viewed across a wide range of frequencies. Freely expanding ejecta is also expected to be present beyond the visible Crab. Here we focus on radio observation of the Crab Nebula. The origin of radio emission is described below.

{\em Radio frequencies}:  The emission in radio frequencies is primarily the non-thermal synchrotron radiation from the shocked pulsar wind; charged particles are accelerated due to the magnetic field of the pulsar. Visually, one can see the expanding network of filaments and fine-scale wispy features (Figure \ref{fig:fig1}). 
The source of most of the thermal emission from the filaments is understood to be the photo-ionization by the hard continuum from the synchrotron PWN emission \citep[see, e.g.,][]{Hester:1996}. In addition, radiative cooling behind the shock driven by the PWN into the freely expanding ejecta also gives rise to radio emission~\cite{Sankrit:1998}.

\subsection{Observational data of the Crab Nebula}
\label{sec:sec3b}
We focus on the data of the Crab Nebula observed at radio frequency by the Very Large Array (VLA) \footnotemark[1]\footnotetext[1]{VLA is operated by the National Radio Astronomy Observatory (NRAO), USA. The National Radio Astronomy Observatory is a facility of the National Science Foundation operated under a cooperative agreement by Associated Universities, Inc.}. We use publicly available radio imaging data of the Crab nebula at 4.76~GHz with an angular resolution of 0.7 arcseconds from the VLA image archive \footnotemark[2]\footnotetext[2]{https://archive.nrao.edu/archive/archiveimage.html}. These data have undergone full calibration and reduction.

The image we analyze is shown in  Figure~\ref{fig:fig1}. The surface brightness in the image is given in units of Jy~beam$^{-1}$. The data has undergone full calibration and reduction. Hence we can directly apply the 2D-MFDFA algorithm to this imaging data and analyze its structural complexity.

\section{Analysis and results}
\label{sec:4}
We apply the 2D-MFDFA algorithm to the radio imaging data of the Crab Nebula and study its multifractal properties. Using the intensity or field value at each pixel, we first calculate the fluctuation functions $F_q$ defined by eqs.~\ref{eq:ffunc} and \ref{eq:Fq0}, and from them, we obtain the generalized Hurst exponents $h(q)$. Then, from $h(q)$, we obtain the derived quantities -  the classical scaling exponents $\tau(q)$, the H$\ddot{o}$lder exponents $\alpha$, the multifractal spectra $f(\alpha)$ and the skewness parameter $\chi$. 

The multifractal analysis using the 2D-MFDFA algorithm was performed using the MATLAB$\_$R2018a platform. The 2D-MFDFA algorithm calculates  $h(q)$ by the method of linear fitting using the inbuilt regression statistics \texttt{regstats} of MATLAB. In particular, it uses the inbuilt routine \texttt{tstat} based on Student's $t$-distribution. The code also calculates the error bars for $h$ as the standard errors from  \texttt{tstat}.

In order to study the spatial scaling behaviors of the multifractal variables over different spatial regions of the Crab Nebula, we repeat the 2D-MFDFA analysis on smaller partitions of the image for both radio and IR data. This is carried out by dividing the images into $N_a\equiv 2^a$ equal-area portions, where $a\ge 0$ is an even integer that determines the number of partitions. We refer to $a$ as the {\em partition scale parameter} (so as not to confuse with the size scale $s$). We use values $a=0,2,4$ since the number of pixels in each partition becomes too small beyond that. The spatial scaling behavior will reveal whether the same morphological rules are followed or not at different parts of the system. It will also enable us to identify the presence of heterogeneous properties at different length scales.

The regions on the top left and bottom right of the image in Figure \ref{fig:fig1} are predominantly noise. Since our calculations require a uniform size of spatial regions, we retain these regions and perform all multifractal analyses. We present the corresponding results in Section~\ref{sec:sec4a} and \ref{sec:sec4b}. In Section~\ref{sec:sec4c}, we analyze the effect of the noise on the multifractal parameters by adopting a masking procedure (described therein the section) on the radio image of Figure \ref{fig:fig1}. We then compare the results for the original (unmasked) and masked radio data.

  
\subsection{Complexity of Crab nebula at the radio frequency of 4.76 GHz}
\label{sec:sec4a}
The Crab Nebula image at the radio frequency of $4.76$ GHz (shown in Fig.~\ref{fig:figm}) consists of $650 \times 740$ number of pixels. 

\begin{figure} 
    \centering   
       \includegraphics[height=14cm,width=8cm]{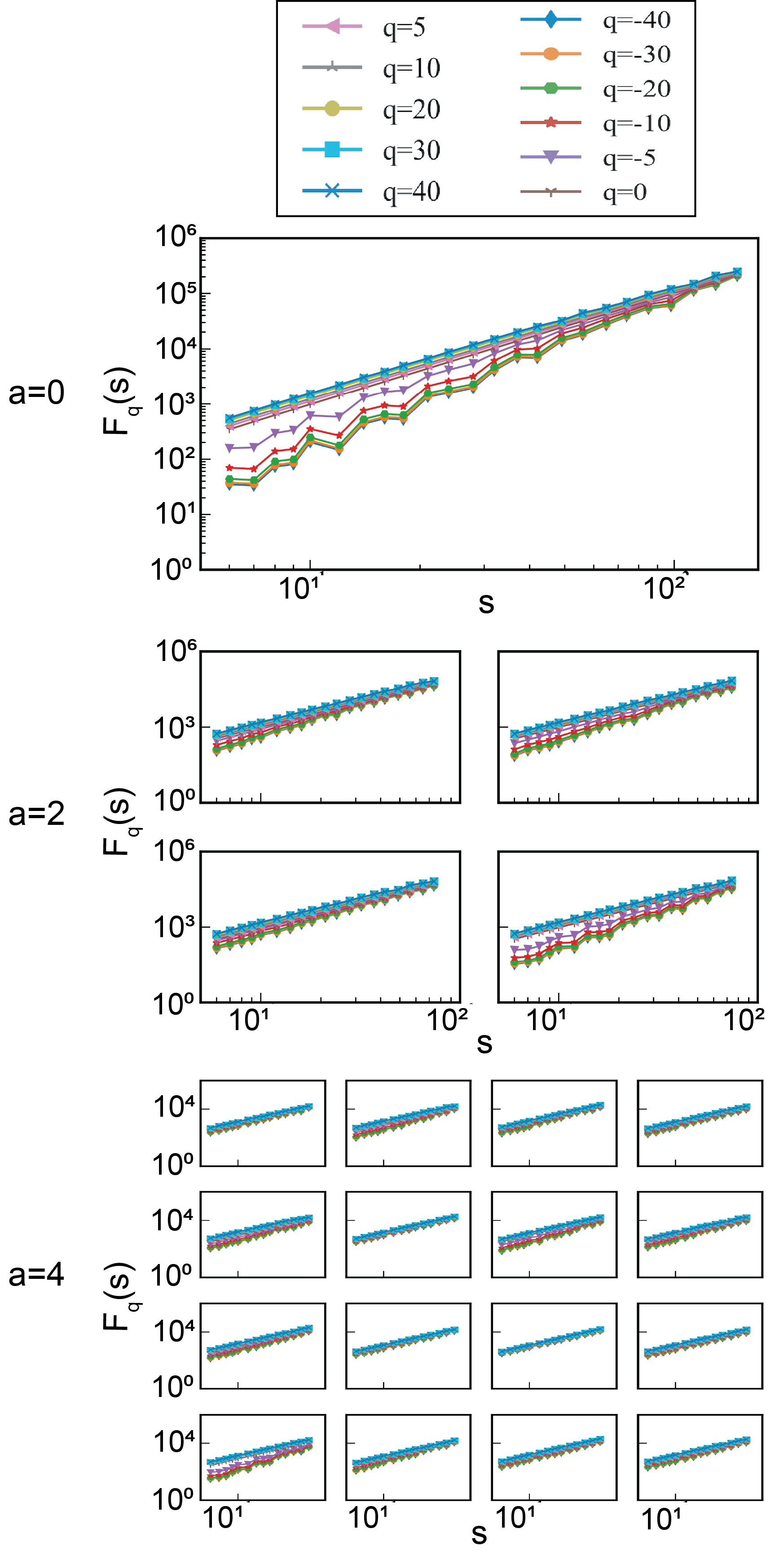}
\caption{ {\em Radio data of Crab Nebula at 4.76 GHz} - The fluctuation function $F_q$ is plotted versus  scale size $s$, for different values of $q$, and different partitions of the radio image corresponding to $a=0,2,4$.} 
 \label{local_radio1}
  \end{figure}

Figure \ref{local_radio1} shows the log-log plots of the fluctuation functions $F_q$ as functions of the scale size $s$, for different partition scales $a=0,2,4$ (indicated against each panel) for different values of the order of statistical moments $q$ (shown by different colors). For all the partition scales $a$, it is observed that $F_q$ has power law dependence on $s$ and the slopes of $F_q$ vary for different $q$. This is a signature of the presence of multiple local fractal behaviors at different local domains $s$, across all $a$. The spread of the curves along $F_q$, i.e. $\Delta F_q$ for small values of $s$ is much larger than the spread in $F_q$ at large values of $s$. This shows that the role of fluctuations is more evident in smaller local patterns/clusters of the system, which provide distinct local behaviors. However, at large values of $s$, this spread $\Delta F_q\rightarrow 0$, where the system converges to a single behavior.

\begin{figure*}  
\centering
    \includegraphics[height=16cm,width=14cm]{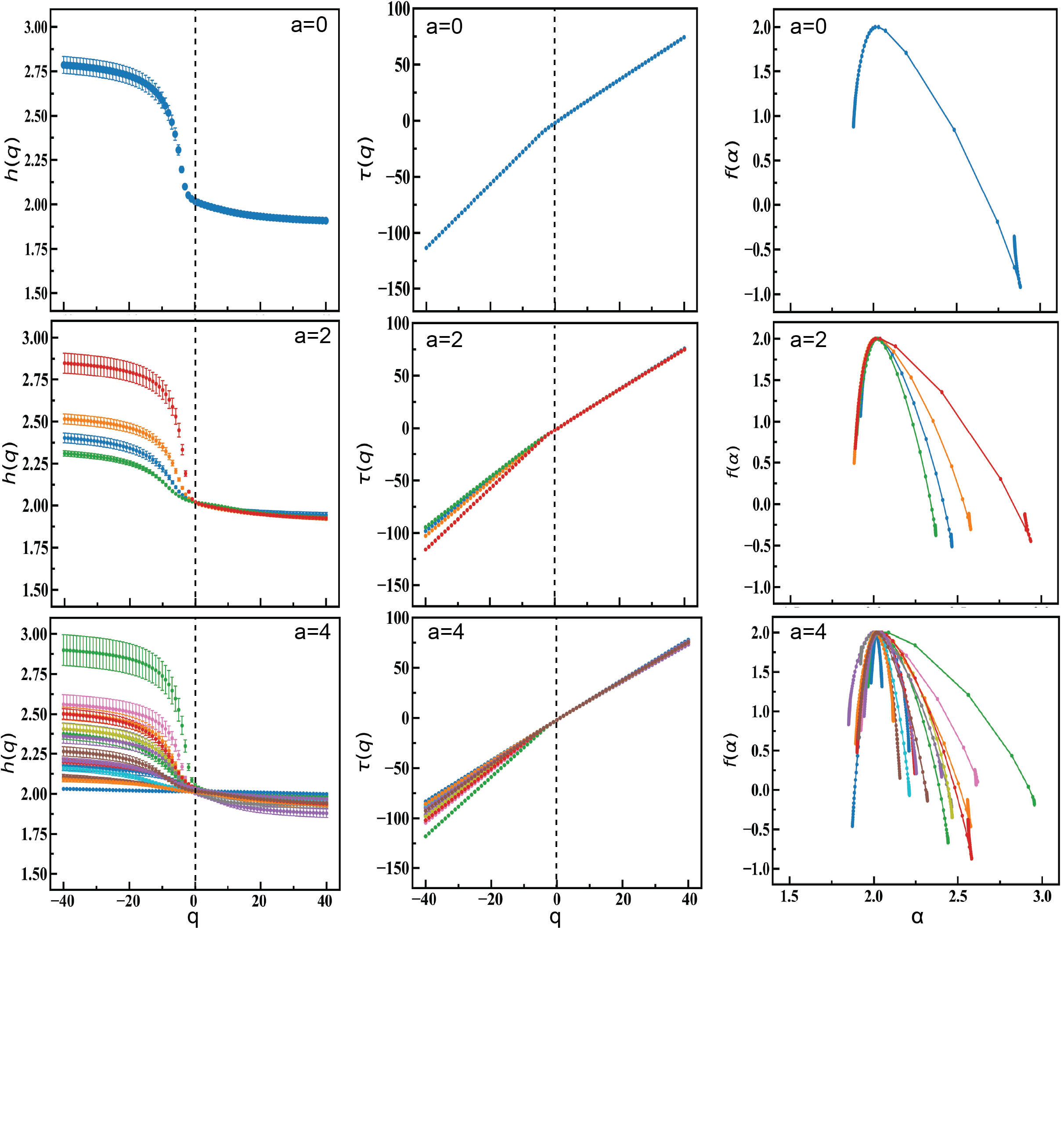} 
   \vspace{-3cm}
    \caption{Multifractal parameters for the radio image of the Crab nebula.   {\em Column 1:} Generalized Hurst exponents $h$ versus $q$ for $a=0,2,4$. 
    {\em Column 2:} The classical scaling exponents $\tau$ versus $q$ for $a=0,2,4$. {\em Column 3:} Multifractal spectra $f$ versus the singular exponents $\alpha$ for $a=0,2,4$. Different colours represent $N_a=2^{a}$ partitions for each value of $a$. The vertical dashed lines in the panels for $h$ and $\tau$ represent the value $q\sim 0 $, where the slopes make a transition. } 
    \label{local_radio2}
\end{figure*}

The slopes of the log-log $F_q$-$s$ plots in Figure \ref{local_radio1} give the values of the generalized Hurst exponent $h$ as a function of $q$, as defined by eq.~\eqref{eq:scaling}. The left column of Figure \ref{local_radio2} shows $h$ versus $q$, ranging from -40 to 40 in steps of one. The cases of  $a=0,2,4$ are shown in the panels from top to bottom. The number of plots (shown in different colors) in each panel corresponds to the number of partitions of the radio image for each $a$. The vertical dashed lines in the panels showing $h(q)$ and $\tau(q)$ correspond to $q\sim 0$. We observe significant dependence of $h$ on $q$ for all $a$, indicating distinct local behaviors at various partition scales $a$. This shows that small and large fluctuations of the emission at the radio frequency have very different scaling natures.  

As described in section~\ref{sec:sec2B2}, for positive (negative) values of $q$, $h$ describes the scaling behavior of the segments with large (small) fluctuations. From the left column of Figure \ref{local_radio2}, we find that the values of $h(q)$ are larger for $q<0$ than that of  $q>0$ for all partition scales $a$. Hence, local behaviors of the patterns/clusters at various $a$ are found to be sensitive to \textit{small fluctuations}. These small fluctuations drive the local behaviors significantly different compared to the large fluctuations in the radio data \cite{Kantelhardt_2002}. The significant spread in $h$ as a function of $q$, i.e. $\Delta h$ at each $a$ shows that scaling laws of the patterns/clusters at various $a$ are distinctly different, which shows a heterogenous distribution of the patterns/clusters in various partitions of the system. Further, since the values of $h(q)$ for $q<0$ are larger than that of $q>0$, the correlated memory in the patterns/clusters/system driven by small fluctuations persists longer as compared to that driven by large fluctuations \cite{RePEc:cwl:cwldpp:1164,Kantelhardt_2002}.

Visually, in the radio image of the Crab Nebula (see Figure~\ref{fig:fig1}) we can discern that in general large fluctuation values track the structurally dominant network of filaments and are correlated over large length scales. In contrast, the smaller fluctuation values track the small-scale wispy structures and are correlated over shorter length scales. Hence, our finding that $h(q>0) < h(q<0)$ is not surprising. It encapsulates the long-range correlations of the filamentary network and the shorter-range correlations, as well as the stronger scale dependence of the wispy structures. Further, we observe scale-dependent heterogeneous patterns in the various partitions, which show different local multifractal scaling laws of the patterns.

In all the plots of $h(q)$, we see that it asymptotes to constant values as $|q|\to \infty$. This implies that small fluctuation values tend towards the same scaling behavior, and so do large fluctuations.   Let $h_{\pm}$ denote the asymptotic values at $q\to\pm\infty$.
We observe that $h_{-}$ varies considerably large across the different partitions for $a\ge 2$, while $h_{+}$ shows relatively less variations such that $h_{\pm}\approx h_{\pm}(a,m)$, where $m$ is the partition index. This means that the structural properties associated with small fluctuations vary considerably large amongst the different partitions of the radio image, while large fluctuations are comparable. Further, since the values of $h_{-}>h_{+}$, the persistence of the patterns/clusters in each partition driven by small fluctuations has longer memory as compared to large fluctuations. We thus expect to see more heterogeneous or dissimilar structures with distinct local scaling laws across the different partitions of the radio image significantly contributed by small fluctuations. This corroborates what we described in the previous paragraph.  We also find the largest variation of $h_{-}$ across the different partitions for $a=4$. A closer look at the values of  $h_{\pm}$ for different partitions of the radio image will be discussed in section~\ref{sec:sec4c}.

In the middle column of Figure \ref{local_radio2}, we plot the classical scaling exponent $\tau$ versus $q$ for different values of $a$. In all the plots, we observe that $\tau$ does not depend on $q$ linearly. Eq.~\eqref{eq:tau} tells us that if $h$ is independent of $q$, then $\tau$ should be a linear function of $q$. But $h$ is dependent on $q$, and it exhibits a step-like transition around $q\sim 0$ (see the left column of Figure \ref{local_radio2}). Therefore, we expect that $\tau$ will also make a transition around $q\sim 0$. This is indeed what we see in the plots of $\tau$, where we observe two linear regimes with a slope change around $q\sim 0$  (indicated by the vertical dashed lines). This value marks the transition of the scaling nature between large and small fluctuations exhibiting multifractal properties.

Lastly, in the right column of Figure \ref{local_radio2}, we plot the multifractal spectra $f(\alpha)$ versus the singularity strength $\alpha$ for different partition scales $a$. Since $h(q)$ is a monotonous function, $\alpha$ is single-valued. We observe that all the curves have their maxima, denoted by $f_{\rm max}$ at roughly $\alpha=\alpha_0\sim 2$. Also, visually it is clear that all $f$ curves are right-skewed around $\alpha_0$, showing a high degree of complexity in the structure of each partition defined by $a$ \cite{stovsic2015multifractal}. Further, the widths of $\alpha$ for a fixed value of $f(\alpha)$ of all the partitions of each $a$ are significantly different, which shows heterogeneous structures in the partitions with different scaling behaviors \cite{shimizu2002multifractal}. The values of the skewness parameters $\chi$ for each $f$ are calculated using eq.~\eqref{eq:skewness}. We will discuss $\chi$ in detail in section~\ref{sec:sec4c}. 

\subsection{Origin of the multifractal nature}
\label{sec:sec4b}
\begin{figure*} 
   \includegraphics[height=5.0cm,width=16cm]{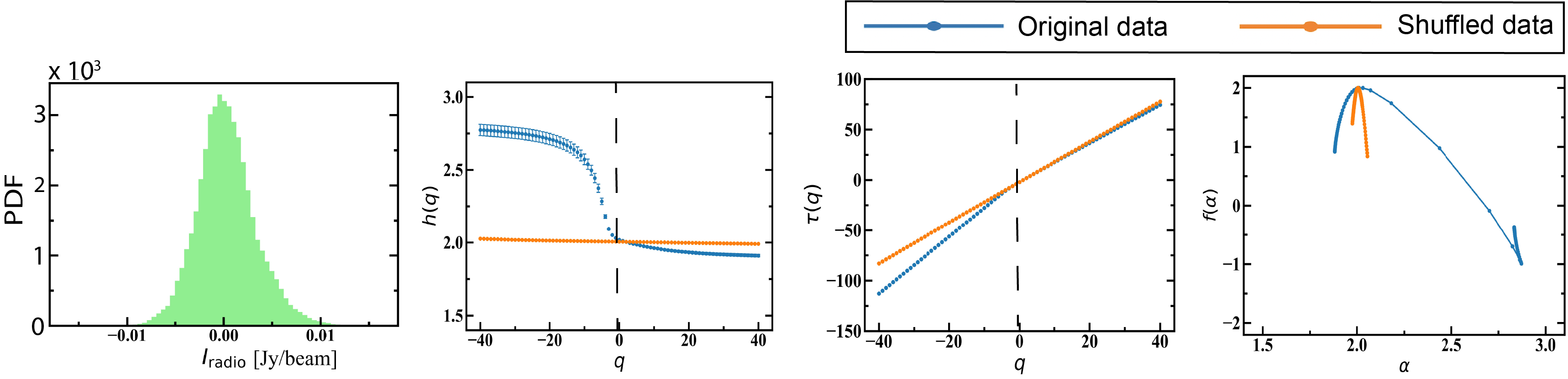} 
    \vspace{-0.3cm}
    \caption{{\em Left:} PDF of the radio data. The $x$-axis is in units of Jy beam$^{-1}$. The other three panels show  $h(q)$ (second from left), $\tau(q)$ (third from left) and $f(\alpha)$ (right) of the radio data for the original and shuffled cases.
    }
    \label{fig:pdf_radio}
\end{figure*}

We now probe the physical origin of the multifractal nature that is obtained for the radio data. Multifractality can arise due to two different physical reasons- (1) distribution-induced multifractality caused by a heavy-tailed probability density function (PDF) of the data, and (2) correlation-induced multifractality caused by differing nature of correlations for small and large fluctuations, such as linear and nonlinear correlations \cite{Kantelhardt_2002,jiang2019multifractal}. One can analyze the origin of multifractality as follows. Shuffling the data destroys all the long-range correlations while the PDF remains unchanged. On the other hand, multifractality due to a heavy-tailed PDF cannot be removed by shuffling \cite{gomes2023origin}. If the multifractality is solely due to correlations, the shuffled data will show monofractality \cite{gomes2023origin}. If the data contains both heavy-tailed PDF and linear/nonlinear correlations, then the shuffled data will show multifractality smaller than that of the original data \cite{barunik2012understanding}. First, in order to isolate the effect of correlation, we randomly shuffle the data to destroy all spatial correlations. We will use superscripts `orig' and `shuf' to indicate quantities obtained from the original and shuffled data, respectively. Suppose $F_q^{orig}$ and $F_q^{shuf}$ are the fluctuation functions corresponding to the original and shuffled data with their respective generalized Hurst exponents $h^{orig}$ and $h^{shuf}$. Then the ratio $\frac{F_q^{orig}}{F_q^{shuf}}\sim s^{h^{corr}}$, where $h^{corr}=h^{orig}-h^{shuf}$. We can analyse the origin of multifractality in the data from $h^{corr}$ as: (i) if $h^{corr}=0$ such that $h^{orig}=h^{shuf}$, then the cause of multifractality is the fatness of the PDF, and (ii) if $h^{corr}=h^{corr}(q)\ne 0$ and $h^{shuf}=h^{shuf}(q)$, then the origin of multifractality is due to both $q$-dependent long-range correlations coming from short and long-range fluctuations as well as a broad PDF in the data, and (iii) if $h^{shuf}= 2$, then the multifractality in the data is due to only long-range correlations 
\cite{Kantelhardt_2002, shang2008detecting}.

In the leftmost panel of Figure \ref{fig:pdf_radio}, we show the PDF for the radio data. We find that it is close to Gaussian. We calculate the multifractal quantities $h(q),\ \tau(q)$ and $f(\alpha)$  for the shuffled data. The second from the left panel of Figure \ref{fig:pdf_radio} shows $h(q)$ for the two cases of original and shuffled data. We indeed find $h^{shuf}(q)$ has weak $q$-dependence with $h^{shuf}(q)\rightarrow {\rm constant} \rightarrow 2$ for the shuffled data. Hence, $h^{corr}\rightarrow h^{orig}(q)-2$. Similarly, we plot  $\tau(q)$ and $f(\alpha)$ for the two cases of original and shuffled data, respectively, in the third from left and right most panels of Figure \ref{fig:pdf_radio}. We previously found that the scaling exponents $\tau(q)$ at all the considered partition scales $a = 0,2,4$ are nonlinear functions of $q$ (see middle panels of Figure \ref{local_radio2}). This non-linearity shows the presence of intrinsic multifractality \cite{zhou2006inverse}. When we shuffled the original field values and repeated the same multifractal analysis, we found that the $\tau(q)$ curve now becomes linear (second from the right panel in Figure \ref{fig:pdf_radio}). This shows that the shuffling process destroys all the intrinsic non-linear correlations.

Since the PDF of the radio data is close to Gaussian, we simulated an uncorrelated Gaussian random field to have a handle on what to expect of the multifractal parameters for uncorrelated Gaussian data (see Appendix~\ref{app:A1}). By repeating the same analysis we performed with Crab Nebula data, we have calculated $h(q),\ \tau(q)$ and $f(\alpha)$ for one realization of an uncorrelated Gaussian random field. We find that for this uncorrelated field, $\tau(q)$ curves show monofractal behavior. From this result in appendix~\ref{app:A1}, we expect the shuffled radio data to behave close to a monofractal and hence to find $h(q)$ to be almost constant with value 2. This is in agreement with what we observe in the panels of column 2 in Figure \ref{fig:pdf_radio}, where we find $\tau(q)$ curves to be linear thereby indicating monofractal behavior. We show the breaking of intrinsic non-linear correlations responsible for intrinsic multifractality by the shuffling process. Therefore we conclude that the multifractality of the Crab nebula radio data originates from long-range non-linear correlations of large and small fluctuations in the data.

\subsection{Heterogeneity in the spatial structures}
\label{sec:sec4c}

\begin{figure}
\includegraphics[height=6cm,width=6cm]{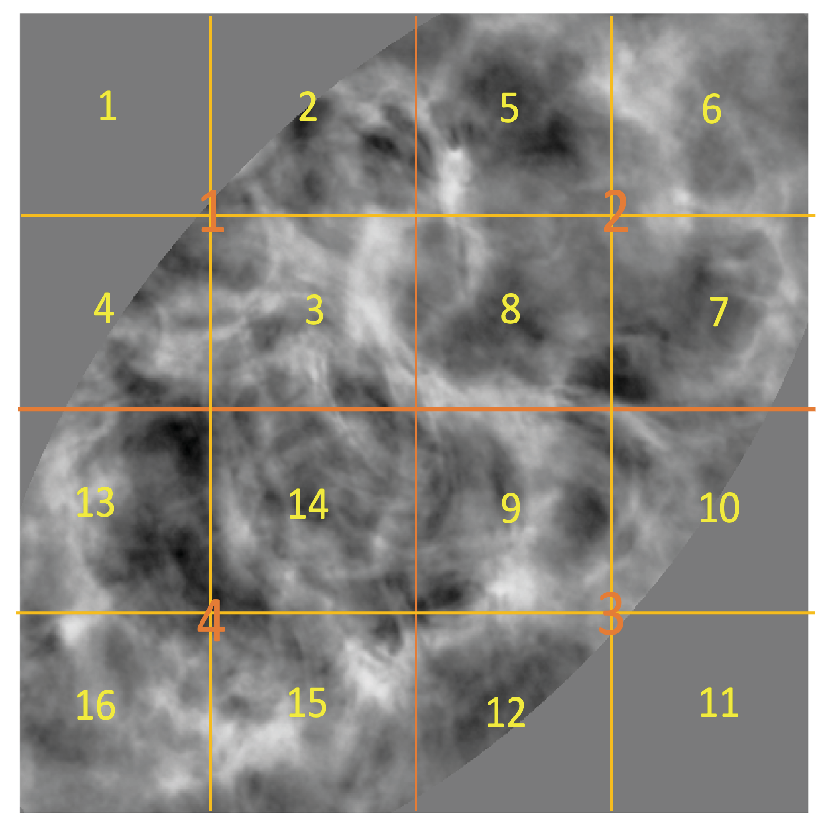}
  \caption{Partitions of the masked VLA image of the Crab Nebula at 4.76 GHz.}
  \label{fig:figm}
  \end{figure}
  
\begin{figure}
\includegraphics[height=6cm,width=9cm]{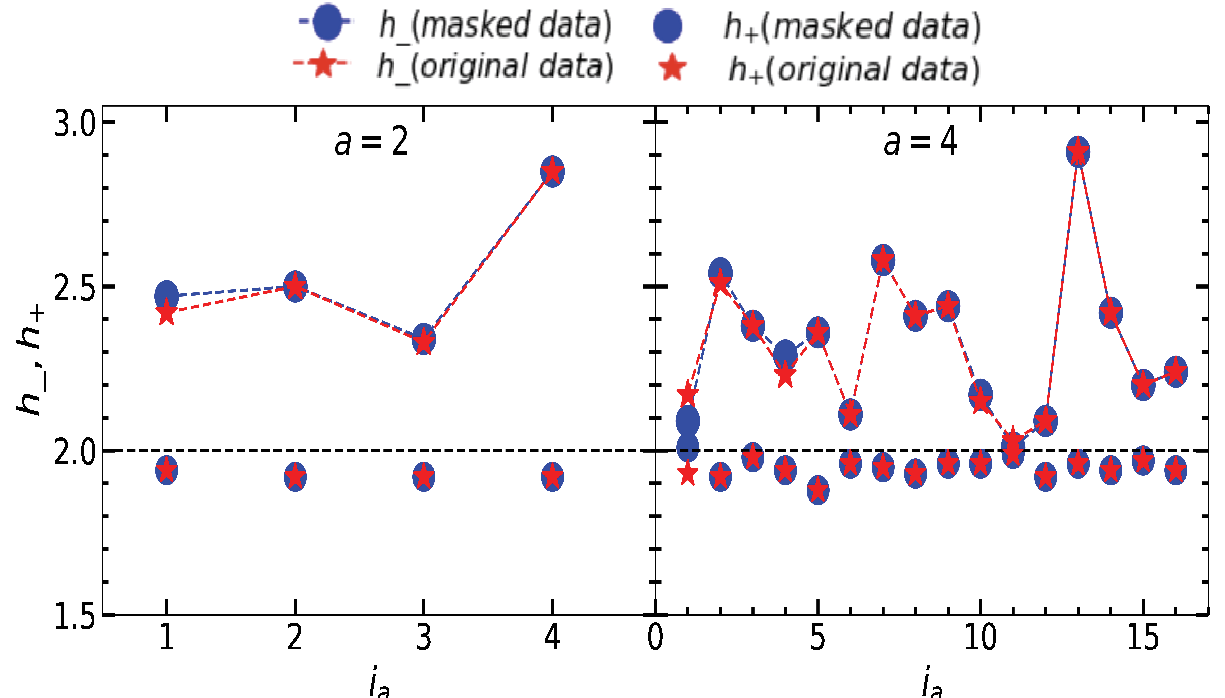}\\
    \caption{Plots of $h_{\pm}$ for each partition indexed by $i_a$ when $a=2,4$ for both masked and original radio image. The black dashed horizontal line shows $h=2$ for reference.}
    \label{fig:fig9}
    \end{figure}

We now study the heterogeneity in the structures of the Crab Nebula observed at the radio frequency of $4.76$GHz by examining integrated quantities obtained from $h$, $\tau$ and $\chi$ as functions of the scale parameter. 

To address the effect of noise present in the top left and bottom right regions of the image in Figure~\ref{fig:fig1}, we repeat the calculations of the quantities  $h$, $\tau$ and $\chi$ after masking these regions. We construct an elliptical mask of appropriate dimensions \footnote{The region is chosen by eye. Minor fluctuations of the size do not have a significant effect on the results.} and assign zero values to the regions outside the ellipse while retaining the original field values inside the ellipse (see Figure \ref{fig:figm}). The partitions in the image are numbered for the cases of $a=2$ (orange) and $a=4$ (yellow). In the latter case, the partitions are numbered in a nested manner so that sequential groups of four are mapped hierarchically to the corresponding larger partition of $a=2$. All calculations are then repeated for the masked image.

We first examine the asymptotic values $h{\pm}$ for the different partitions indexed by $i_a$ (for a given partition scale $a$). The plots of $h{\pm}$ versus $i_a$ are shown in Figure~\ref{fig:fig9} for $a=2$ (left panel) and $a=4$ (right panel). Red stars indicate values for the unmasked original image, while blue indicates values for the masked images. For $a=2$, there is an increase in $h\_$ for $i_a=1$ and a slight increase in $h\_$ for $i_a=3$ after masking. Similarly, for $a=4$, there is a  change in $h\_$ for $i_a=1,2,4$ and a slight change in $i_a=10,11,12$ after masking. 
These changes are because the mask artificially introduces two distinct regions in each affected region - one that belongs to the Crab nebula and another that has predominantly zero values.    

From the bottom panels of  Figure~\ref{fig:fig9}, we can see that for $a=2$, the radio image (both original and masked) has larger values of $h_-$ and smaller values of $h_+$. This captures the wider variation of scaling behavior between large and small fluctuation values of the radio data, already discussed earlier.  This is corroborated by visual inspection of the image. 
If we closely look at the smaller partition scale structures ($a=2$) of the Crab nebula observed at radio (top panel of Figure~\ref{fig:fig9}), we can easily see that the radio image has more heterogenous structures across the partitions in the form of filaments and wisps. Moreover, the filamentary network of the radio image has longer-range correlations. This can be explained by the fact that the materials that emit synchrotron radiation at radio frequencies, which are the charged particles of the supernova ejecta and pulsar wind, are distributed over larger spatial scales. At $a=4$, $h_{-}$ for the radio data varies randomly. This implies that the structures across the different parts of the Crab Nebula are highly heterogeneous when observed at radio frequency. These highly heterogeneous structures at various length scales make the Crab Nebula a structurally complex physical system.   

Next, we focus on the values of $h$ in the regime around $q=0$, where it exhibits a transition. Let the average of $h(q)$ over $q$ in a suitable regime $-q_c<q<q_c$, with $q_c$ is a positive variable, be given by
\begin{equation}
\bar h \equiv 
\frac{1}{\int_{-q_c}^{q_c} {\rm d}q} \int_{-q_c}^{q_c} {\rm d}q\, h(q). 
\label{eq:hbar}
\end{equation}
In the limit $q_c\to \infty$,   $\bar h$ will be dominated by the asymptotic values, and we simply get $\bar h \sim (h_{-}+h_{+})/2$. 
In this limit, we lose the information of the variation or shape of $h(q)$ in the intermediate $q$ values. So we focus on finite $q_c$ and choose $q_c=40$ as the value of $q$, where $h$ becomes effectively constant for each partition scale $a$. This is akin to weighted averaging, where the weight kernel is the top hat function centered at $q=0$. Mathematically there is no
restriction on taking large $|q|$. Since $q$ takes integer values here, we need the discretized version of eq.~\ref{eq:hbar}. Let $i_q=1,\ldots,N_q$ index the values of $q$, where $N_q=81$ denotes the total number of $q$ values considered. Let $i_a=1,\ldots,N_a$ index each partition of the image for a given $a$. Then for each $i_a$, the discrete form of eq.~\ref{eq:hbar} is
\begin{equation}
\bar{h}_{i_a} \equiv \frac{1}{N_q}\sum_{i_q=1}^{N_q}  h_{i_a}(q).
\label{eq:avg1}
\end{equation}
 
For $\tau$, there are no asymptotically constant values. Hence, there is no natural cut-off value of $q_c$ to define the range $-q_c<q<q_c$ over which to average $\tau$. We will choose the same $q_c$ that is defined for $h$. So we have   
\begin{equation}
\bar{\tau}_{i_a} \equiv \frac{1}{N_q}\sum_{i_q=1}^{N_q} \tau_{i_a}(q).
\label{eq:avg1}
\end{equation}

Let $\chi_{i_a}$ denote the skewness parameter $\chi$ in each partition $i_a$. Let $\bar{X}_{i_a}$ denote either of the three quantities - $\bar{h}_{i_a}$, $\bar{\tau}_{i_a}$ or  $\chi_{i_a}$. Then we average them over all the partitions for each $a$, as
\begin{equation}
\langle X \rangle \equiv \frac{1}{N_a}\sum_{i_a=1}^{N_a} \bar{X}_{i_a}.
\label{eq:avg2}
\end{equation}

\begin{figure}  
  \vspace{-.3cm}
 \includegraphics[height=12cm,width=5cm,angle=0]{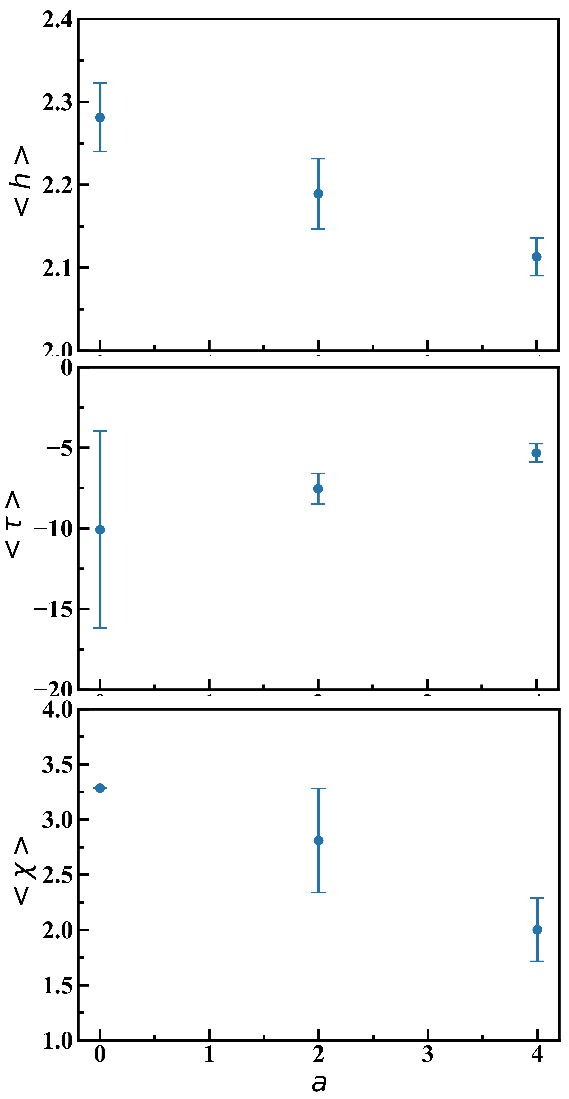} 
 \vspace{-0.3cm}
    \caption{Scaling with respect to $a$ of the averaged quantities $\langle h\rangle$ (top panel), $\langle \tau\rangle$ (middle panel) and $\langle \chi\rangle$ (bottom panel) for radio data.}
    \label{fig:fig10}
    \end{figure}
Figure~\ref{fig:fig10} shows $\langle h \rangle$ (top panel), $\langle \tau \rangle$ (middle panel) and $\langle \chi \rangle$ (bottom panel) for the radio image as functions of partition scales $a$. The error bars shown are the standard errors.  
We observe that $\langle h\rangle$ varies with $a$ for the radio data. This result correlates with the value of $(h_-+h_+)/2$ seen in Figure~\ref{fig:fig9}, with the larger contributions coming from $h_-$.  This implies that the radio image has stronger small-scale scaling behaviors, showing the presence of fine-scale structures with well-defined local scaling laws in the radio data. 

Further, let $h_0$ denote the largest value of $\la h\ra$ chosen from the different $a$'s. Let us normalize $\la h\ra$ by $h_0$ and denote 
\begin{equation}
\langle h\rangle_{\rm norm}\equiv \langle h\rangle/h_0.
\end{equation}
If $\langle h \rangle_{\rm norm} \in (0.5,1)$, then this implies a strong positive correlation in the system \cite{Kantelhardt_2002}, with long memory of persistence in the states of the system \cite{RePEc:cwl:cwldpp:1164, kantelhardt2006long}. From the top panel of Figure \ref{fig:fig10}, for the radio data, $h_0=2.28$, which is the value of $\la h\ra$ at $a=0$. Using this, we find that $\langle h\rangle_{\rm norm}$ lies in the range  $[0.92,1]$ for all $a$ for the radio data. This value is greater than 0.5 and close to unity. This implies strong positive correlations~\cite{Kantelhardt_2002} in the Crab Nebula structural system. Further, the decrease in $\langle h\rangle$ as $a$ increases show that finer structures in the smaller partitions of the system lose their correlations and local structural memories, which is expected. In the asymptotic limit $\displaystyle\lim_{a\rightarrow\infty}|\langle h\rangle|\rightarrow 2$, it shows memoryless structures whose underlying processes are Brownian motions (see Appendix~\ref{app:A1}) \cite{RePEc:cwl:cwldpp:1164,Kantelhardt_2002, kantelhardt2006long}. Again, the curve of $\langle h\rangle$ approximately follows a power-law behavior with $a$ as $\langle h\rangle\sim a^{-\gamma}$, where $\gamma$ is the power-law exponent. This shows a hierarchical organization of the heterogenous structures in the Crab Nebula structural system \cite{s1992fractal,ravasz2003hierarchical,jones2005scaling}. This behavior is similar to Carpenter's law as predicted in astrophysical data analyses \cite{de1971large, carpenter1931cluster,jones2005scaling}.

The middle panel of Figure \ref{fig:fig10} shows that the average classical scaling exponent $\langle \tau\rangle$ also varies with $a$, which again implies heterogeneity in the structure of the underlying system. From the bottom panel of Figure \ref{fig:fig10}, $\langle \chi \rangle >2$ for all $a$, which implies the richness of complexity of the Crab nebula at radio frequency. This shows the dominance of scaling of small fluctuations or the presence of fine structures in the Crab Nebula. The random distribution of the $\langle \chi \rangle$ parameter also implies a heterogeneous property of the structure visible in the Crab Nebula. This can be accounted for by the non-equilibrium dynamics of the underlying system.
\section{Conclusions}
\label{sec:5}
The Crab Nebula is one of the most studied astrophysical systems across the entire electromagnetic spectrum from radio wavelengths to ultrahigh-energy $\gamma$-rays \cite{hess2020resolving,bucciantini2023simultaneous}. In astrophysics, several recent studies confirm the complexity of the Crab Nebula’s morphology across the different bands of the spectrum \cite{yeung2021studies,liu2021pev,chastenet2022sofia,wootten2022dense,ponomaryov2023origin,meyer2023plerionic}. In this paper, we offer a fresh perspective on the complex morphology of Crab Nebula using multifractal analysis from the vantage point of complex systems science.

We have investigated the structural complexity of the Crab Nebula at the radio frequency of $4.76$ GHz using publicly available imaging data. In particular, we have analyzed the data using the 2D-MFDFA approach. We have investigated the structural system at various length scales for local scaling behaviors. Our results confirm that the structure of the Crab Nebula has a rich complexity characterized by a multifractal nature. We have studied the origin of multifractality in the radio data. We have found that the multifractal property of the radio data arises from the difference in the nature of long-range correlations (and consequent scaling with the scale parameter $s$) of large and small field values in the data. 

Our results show a strong positive correlation in the radio frequency. The long-range correlations of large and small values in the data can be traced back to the materials that emit at this frequency. The variations of the multifractal parameters, namely the generalized Hurst exponents $h$ calculated from the fluctuation functions $F_q$, the classical scaling exponents $\tau$ and the skewness parameter $\chi$ across varying length scales show heterogeneity in the structure across different parts of the visible Crab Nebula. This can be accounted to the Crab Nebula being a non-equilibrium, highly dynamic structural system.

What we have done in this paper is essentially quantified how the physics of the evolution of the Crab Nebula as a supernova remnant manifests in the language of complex systems. The 2D-MFDFA approach we have used in the study is found to be a good method to characterize the complexity of an astrophysical system with complex morphological features. Our multifractal results inform us that the rich fine structures in the Crab Nebula ruled by local scaling behaviors show self-organization at different length scales of the structural system. Long-range correlations manifest as self-similar fractals and are identified as signatures of self-organized criticality \cite{bak1988self,bak1989physics}. Our results thus deepen our understanding of the multi-scale physics that is operating in the Crab Nebula. An example of a complex system with multifractal characteristics similar to the Crab Nebula is the human brain. Multifractality captures the heterogeneous and multiscale interaction rules in the brain networks \cite{yang2020controlling}; for instance, the community structures at different structural levels in the human brain connectome \cite{xue2017reliable} ruled by scaling behaviors that are signatures of self-organization \cite{singh2016scaling}. 

This paper represents a proof of concept of the insights that can be obtained from viewing the Crab Nebula as a complex system. However, it has some limitations. First, our analysis considers the Crab as a two-dimensional object and ignores its three-dimensional structure. It will be interesting to study the full 3D structure using reconstructions such as those given by~\cite{Martin:2021}. Secondly, the Crab Nebula is a rapidly evolving system on astrophysical time scales. The data that we analyze here are observations taken at specific times, and we do not address the question of the time evolution of the properties related to complexity. 
In order to extract the full potential of the method, our work can be extended in the following directions. 
Given the wealth of data available for this astrophysical object, an analysis across the full range of frequencies will be valuable and will be carried out in the future. It opens up a slew of questions that will enhance our understanding of supernova remnants, and we plan to carry out a systematic investigation in the near future. 

The methodology used in this paper can also be used to understand other astrophysical complex systems such as the planetary nebula NGC 1514, whose complex kinematics with multiple structures was studied \cite{aller2021morpho}. As pointed out in the review paper \cite{olmi2023dawes}, pulsar wind nebulae, f which the Crab nebula is a prototype, show a variety of properties and morphologies at different evolutionary ages. It will be interesting to extend the present analysis to study the multifractal properties of a wide variety of pulsar wind nebulae and look for their universal properties.

\section*{Authors' contributions}
{\noindent}The conceptualization of the present work is done by ALC, PC, and RKBS. PC and FR processed the data of the Crab Nebula for multifractal analysis. ALC and FR performed the multifractal analysis. ALC, PC, and FR prepared the associated figures. PK vetted the observed data. All authors wrote, discussed, and approved the final manuscript.

\section*{Acknowledgements}  
\noindent We thank Prof. Wei-Xing Zhou for providing the 2D-MFDFA code for our multifractal analysis. ALC acknowledges financial support from the Indian Institute of Astrophysics (Bengaluru), the Department of Science and Technology Government of India (Order No: \texttt{DST/INSPIRE Fellowship/2018/IF180043}), the APCTP (YST program) through the Science and Technology Promotion Fund and Lottery Fund of the Korean Government and the Korean Local governments-Gyeongsangbuk-do Province and Pohang city. The work of PC~is supported by the Science and Engineering Research Board of the Department of Science and Technology, India, under the \texttt{MATRICS} scheme, bearing project reference no \texttt{MTR/2018/000896}. RKBS acknowledges funding from the Science and Engineering Research Board of the Department of Science and Technology, Government of India, under the \texttt{MATRICS} scheme, bearing project reference no \texttt{MTR/2021/000766}. PK acknowledges funding from the Department of Atomic Energy, Government of India, under the project {\tt 12-R\&D-TFR-5.02-0700}. 

\section*{Conflict of Interests }
\noindent The author(s) has no conflicts to disclose.

\section*{Data Availability Statement}
\noindent The data that support the findings of this study are available from the corresponding author upon reasonable request.

\appendix
\section{Multifractal parameters for uncorrelated Gaussian random data on two dimensions}
\label{app:A1}

To serve as a standard reference with which to compare the multifractal parameters of the Crab Nebula at different frequencies it is useful to know the expected behavior of the parameters for an uncorrelated Gaussian random field. For this purpose, we simulated one realization of such a field having zero mean and unit variance on a $512\times 512$  grid. Then we calculated the multifractal parameters using the same 2D-MFDFA algorithm that is used for the Crab Nebula data. 

\begin{figure*}
\includegraphics[height=4cm,width=16cm]{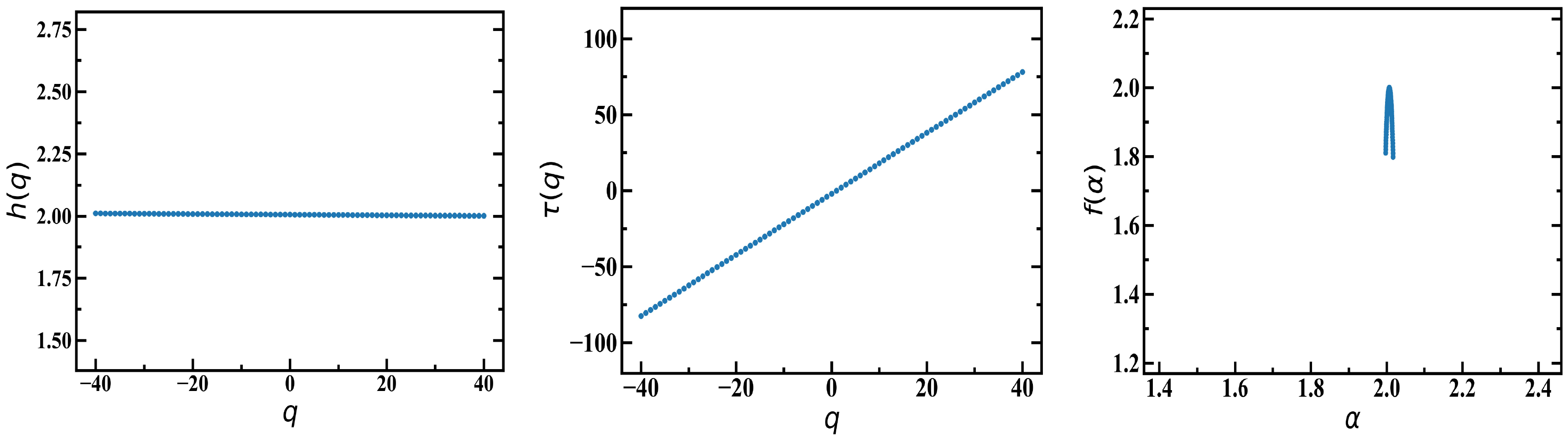}
    \caption{Multifractal parameters $h(q)$ (left), $\tau(q)$ (middle) and $f(\alpha)$ (right) for one realization of an uncorrelated two-dimensional Gaussian random field having zero mean and unit variance.}
    \label{fig:gaussian}
\end{figure*}

The resulting plots of $h(q)$, $\tau(q)$ and $f(\alpha)$ are shown in Figure \ref{fig:gaussian}. From the left panel, we see that $h(q)$ is constant $\sim 2$, which implies a monofractal behavior as expected from the uncorrelated and Gaussian natures of the field. The linearity of the plot for $\tau(q)$ (middle),  and narrow range for $f(\alpha)$ (right), follow immediately from the plot of $h(q)$. The very slight variation of $\alpha$ on the $x-$axis of the third panel arises from the statistical fluctuation of the data since we have used only one realization of the Gaussian random field.

\bibliographystyle{refstyl}
\bibliography{references}
\end{document}